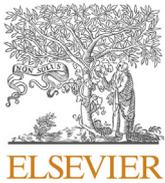
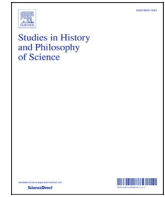

# Functionalising the wavefunction

Lorenzo Lorenzetti

*University of Bristol, Department of Philosophy, Cotham House, BS6 6JL, UK*

## ABSTRACT

Functionalism is the view that being *x* is to play the role of *x*. This paper defends a functionalist account of three-dimensional entities in the context of Wave Function Realism (WFR), that can explain in details how we can recover three-dimensional entities out of the wavefunction. In particular, the essay advocates for a novel version of WFR in terms of a functional reductionist approach in the style of David Lewis. This account entails reduction of the upper entities to the bottom ones, when the latter behave appropriately. As applied to WFR, it shows how the wavefunction can turn out to be identical to three-dimensional objects, provided certain conditions. The first major goal of the paper is thus to put forward an improved and more rigorous version of WFR, which dissolves several extant issues about the theory, and can serve as a starting point for the future literature about the topic. Moreover, the second major goal of the article is to take WFR as a case study to demonstrate the pros of functional reductionism, especially in the form defended here, thereby helping to bring this view back in the philosophy of science debate. The positive upshots of this paper suggest a possible application of functional reductionism also to other contexts.

## 1. Introduction

### 1.1. The starting point

Wave Function Realism (WFR) is one of the most prominent accounts concerning the ontology of quantum mechanics. The central core of WFR, as defended by David Albert (1996, 2013, 2015) and Alyssa Ney (2012, 2015, 2020, 2021a, 2021b),[1] is the following:

**Core WFR:** The quantum wavefunction represents a field living in a 3N-dimensional configuration space.[2] This high-dimensional field – alone, or together with a single point particle in the case of Bohmian mechanics – is everything there is fundamentally, and configuration space should be taken as the fundamental space of our universe. The manifest three-dimensional world is, thus, not fundamental.

On top of this, Albert (1996, 2015) and (the early) Ney (2012),[3] have argued that, even if we endorse WFR, we can still make sense of the existence of three-dimensional (from now onward, simply '3D') entities if we embrace the following functionalist thesis:

(i) As long as we characterize 3D objects in terms of their functional role; and,

(ii) As long as there is something within the fundamental 3N-dimensional ontology that can play the role which is associated to 3D objects; then,

(iii) 3D objects exist.

What does it mean to "play the role"? And how should we cash out exactly the notion of functional role? Albert suggests that functional roles should be interpreted in a nomological sense. In our case, to apply a functional understanding of 3D objects means to characterize them in terms of the dynamical equations of classical mechanics. Ultimately, to be 3D is to behave according to those laws. Correspondingly, if we want to say that there is something in the fundamental ontology that plays the role associated to 3D objects, we need to show that there is something within the former ontology which behaves according to the same dynamical equations that govern 3D objects. I refer to the combination of statements (i)-(iii) above and the nomological understanding of functional roles as **Minimal Functionalism** about WFR.

Building on Minimal Functionalism, Albert (2015) has also provided a sketch of the physics underling the functionalist approach to WFR, based on certain formal similarities between Hamiltonians in 3D and high-dimensional spaces, and has argued in favour of a specific metaphysics for WFR which dubs the 3D ontology as 'emergent'. I will call this package **Primitive Ontology Functionalism**.

---

In contrast to him – but sharing the appeal to Minimal Functionalism – this paper defends an alternative functionalist approach to WFR. In particular, the paper puts forward a novel functionalist approach and confronts it with Albert's Primitive Ontology Functionalism, to show its advantages. This account is a form of **Functional Reductionism**, along the lines of Lewis (1970, 1972) and Butterfield and Gomes (2020a, 2020b).[4] According to this view we can reduce low-dimensional entities to high-dimensional ones when and where the 'bottom' entities behave as the 'upper' ones, to a sufficient degree of approximation.

*1.2. Why funtional reductionism about WFR matters*

The goal of this operation is to provide a metaphysically improved and conceptually and formally clearer version of WFR, which can be taken as a starting point for the future debate about WFR. In doing so, I thereby prove that functional reductionism is alive and kicking, and should be taken seriously by philosophers of science in general, especially in the Lewisian form presented here which has been recently revived by Butterfield and Gomes. In other words, I show both how the debate about WFR can benefit from functional reduction, and – at the same time – how we can learn important lessons about the utility of functional reduction by analysing certain issues that have been discussed in relation to WFR. This can arguably have important implications also for other contexts in the philosophy of science where functionalism has been discussed. Thus, WFR can be taken here also as a *case study*, and for this reason the present discussion will be useful also for those who are sceptics about WFR.[5] In fact, I want to suggest that the advantages of adopting functional reductionism that I am going to sum up in a moment can be expected to be extendible also to other contexts in the philosophy of science where functionalism is discussed, such as the debate about spacetime emergence in quantum gravity (see Lam and Wüthrich (2018; 2020)) and the discussion about emergence within Everettian quantum mechanics (see Wallace (2012)). Furthermore, functionalism has been recently discussed also in connection to general relativity (Knox (2014, 2019)) and thermodynamics (Robertson (forth.)). More on this in section 6.

Talking about WFR *per se*, the more specific reasons why the present functional reductionist account of WFR should be regarded as an improvement to the current alternatives are the following. **First**, it shows in a precise way how functionalism works within actual physical situations. Being more specific about the mathematical details underlying the functionalist approach makes the view less abstract, thereby dissolving some ambiguities and philosophical puzzles (discussed for instance by Ney (2021b)) linked to Albert's functionalist account of WFR, which arguably arise due to the lack of mathematical details. **Second**, by providing a more formal characterization of the Minimal Functionalist thesis, we can get a proper account concerning the ontology of WFR and especially about the exact relation between the high-dimensional and the low-dimensional ontologies, and about the ontological status of the latter one. More precisely, by formalizing the functionalist schema and showing how it entails functional reductionism, we are able to replace the generic talk about the 'emergence' of the 3D entities from the wavefunction with a more rigorous account about the nature of the 3D ontology, expressed in terms of reduction and identity relations. Indeed, under functional reductionism, wavefunction states turn out to be identical to configurations of classical 3D particles, when they behave appropriately – that is, exactly when/where we need to have 3D entities, i.e. in those contexts in scientific practise in which we treat quantum systems *as* (approximate) classical systems. **Third**, relatedly, this refined version of WFR can help closing the possible explanatory or ontological gaps between the high and the low dimensional ontologies (thanks to functionally-induced identity relations). I show that closing the gap allows us to easily tackle some metaphysical issues which has been widely discussed over the last years concerning the alleged emergence of the 3D world within WFR. **Fourth**, and finally, the reduction of 3D entities to the wavefunction in the relevant situations where the latter can play the appropriate role associated with 3D objects, gives us an ontology that combines the virtues of WFR and the conceptual clarity of a 3D ontology. In other words, the fundamental ontology of non-relativistic quantum mechanics is the one that is predicated by the defenders of WFR, i.e. a field in configuration space, and is thereby both separable and local, but that same ontology is – *in the relevant contexts* – simply 3D, and thus it is as conceptually clear as a familiar fundamentally 3D one. Thus, functional reductionism allows us to demonstrate that WFR is not that different from other more 'intuitive' ontological accounts of quantum mechanics.

I conclude this introduction with an outline of the paper. Section 2 unpacks and reviews the thesis of Core WFR. Then, section 3 presents my functional reductionist account. In particular, section 3.1 provides some physically compelling examples to show how a wavefunction can behave approximately three-dimensionally *in the appropriate situation* (based on semiclassical features, especially on Ehrenfest's theorem), in order to show in a physically perspicuous way how the functional realization of 3D objects should work. Then, drawing on the previous section, section 3.2 provides a more formalised version of Minimal Functionalism – using the tools provided by David Lewis's (1970) functionalist framework – thereby showing that a particularly sensible way to cash out Minimal Functionalism is in term of functional reduction. I then apply that framework to the case studied in section 3.1. Section 3.3 elaborates further some crucial aspects of this functional reductionist view. Section 4 presents Albert's Primitive Ontology Functionalism, and shows that the view defended here fares better concerning the details about the physics underlying the functionalist approach. Section 5 shows how the account defended here can address in a natural way also several conceptual and metaphysical problems which are left opened by Albert's functionalist view, and which are related both to his peculiar account and to Minimal Functionalism in general. Section 6 discusses the broader upshots of the present discussion.

**2. Wave Function Realism in a nutshell**

In quantum mechanics, physical states can be represented by wavefunctions, which are functions from points in space to complex numbers. In particular, they assign a pair of numbers – amplitudes and phases[6] – to points in the *configuration space*. For instance, the position state of a particle is represented by the value of this function at one particular point in the configuration space. Indeed, configuration space is a crucial feature in quantum mechanics. To see this, consider first the case of classical mechanics. Here, a system of (say) three particles at a certain instant $t = 0$ can be represented via three triplets of Cartesian coordinates. However, the same system can be represented also via a single 9-tuple of coordinates in the $3N$-dimensional configuration space of the system (which is 9D for this system). In this space, the ensemble is denoted by a set of coordinates that picks out a single point.[7] In classical mechanics, the configuration space representation of the system is merely interpreted as a useful alternative representation, whereas the real space is the ordinary 3D one. Differently, in quantum mechanics, on pain of losing important physical information, the configuration space representation for (position) quantum wavefunctions is ineliminable, due to the phenomenon of quantum entanglement.[8]

---

[4] See also Kim (1998, 2005). Notice that this view is crucially different from the 'Wave-functionalism' defended by Allori (2021). Her proposal provides a functionalist analysis of the concept of wavefunction itself, in order to give an account about the physical meaning of the wavefunction. On the other hand, the approach defended here takes stock of the WFR account of the wavefunction (an alternative to Wave-functionalism) and endorses a functionalist account of the 3D entities, to show how the wavefunction can play their roles.

[5] WFR – here inteded as Core WFR – is an account that has arguably several technical limits as an ontological interpretation of quantum mechanics, as crucially highlighted for example by Wallace (see Wallace et al. (Wallace & Timpson, 2010), Wallace (2017, 2020), and responses by Schroeren, 2022 and Ney (2021b)).

[6] I the following I shall ignore phase and focus only on amplitude, as it is customary in the literature.

[7] More precisely, classical mechanics specifies both position and velocity for every particle, so we have three sextuplets of values in the 3D space. Correspondingly, when moving to the high-dimensional space, we should move to the phase space. I focus here just on the position representation for simplicity.

I can now present the central tenet of WFR. I said that the wavefunction assigns numbers to points in space. WFR takes the further ontological step of interpreting the wavefunction as representing a *physical field*. The term 'physical' here denotes that the wavefunction is not a mere mathematical device, but refers to a real object, analogous e.g. to the Maxwellian electromagnetic field (which is not complex, though). However, since the wavefunction assigns numbers to points in the configuration space, this field does not occupy the ordinary 3D space, but it inhabits the 3*N*-dimensional configuration space.[9] On top of that, WFR maintains that this very high-dimensional space is the fundamental physical space of the world.[10] Thus, WFR is mainly a thesis about the nature of wavefunction – i.e. about *which kind of entity* the wavefunction is – and about the dimensionality of the fundamental space of the world. To be more precise, the fundamental ontology of the theory is simply constituted by spatial points (not the ordinary spacetime points, but the points in configuration space) instantiating intrinsic properties.[11]

This theory can be contrasted with another influential approach to the ontology of non-relativistic quantum mechanics, i.e. *the primitive ontology approach* (Maudlin, 2007; Allori et al., 2008). According to this view, the ontology of quantum mechanics is fundamentally located in 3D space – which is the fundamental space – whereas the wavefunction should be interpreted as 'operating' on that ontology. In this sense, the wavefunction can be taken e.g. as a law of nature, a disposition, and so on.[12] Each version of quantum mechanics (e.g. GRW theory, Bohmian mechanics) can be tied to a specific primitive ontology – for instance, made of 3D matter density fields in the case of GRW theory.

The main reasons why WFR has been defended in the literature are the facts that it is an ontological reading of (non-relativistic) quantum mechanics which follows straightforwardly from the mathematics of the theory,[13] and that WFR allows us to have an ontology for quantum mechanics which is both fundamentally *separable* and *local*. A state can be said to be non-separable if it is not wholly fixed by the states of the subsystems which constitute it, while non-locality is said to take place when we have instantaneous action at a distance.[14] These puzzling features – which can be addressed by WFR – are often considered as consequences of quantum entanglement (see Ismael (2020); Ney (2021b)). Take for instance two particles in a singlet state, in 3D space. It can be argued that this state is not separable, and that measurement on one party of the singlet state will instantaneously influence the state of the other party, independently from their distance, thereby entailing non-locality. It is an open question whether non-locality and non-separability are really negative features, one which I will not address here. However, proponents of WFR argue that this is the only account of quantum mechanics that can provide a fundamental ontology which is both local[15] and separable, and they regard this as a pivotal virtue of their account: "The state of the total field [i.e. the wavefunction in configuration space] is determined by its state at each point and there is no action at a distance" (Ney (2021b, p. 18).[16] In fact – as the wavefunction realist claims – what we see as numerically distinct (entangled) particles in 3D space, are really just a manifestation of a single more fundamental entity in the high-dimensional space.

On a side note, let's notice that with respect to the particular versions of quantum mechanics that one can endorse (e.g. Everettian quantum mechanics, etc.), different sorts of WFR can be distinguished, which entail distinct accounts about the fundamental ontology of the world (cf. Albert (2013), p. 54). Roughly put, depending on which theory one defends, one will claim that the high-dimensional wavefunction is everything there exists fundamentally, or will claim that we have to postulate an additional piece of ontology in the 3N-dimensional space. However, I will not go further into this issue here. Rather, I want to focus just on the general picture that every specification of WFR will share, since every version of WFR follows the same pattern: it postulates a fundamental ontology in the 3*N*-dimensional space and it claims that this fundamental ontology "determines" (in a sense to be clarified) a 3D-located ontology.

## 3. Functional reductionism and wave function realism

Section 2 introduced the basics of WFR. This section draws on that and on Minimal Functionalism to build a complete functionalist account of WFR, which can provide a detailed picture about the relation between the 3D and the high-dimensional levels in WFR, and improve the existing literature.

### 3.1. Functionalisation in action

Minimal Functionalism introduces a *two-steps strategy* to recover 3D entities. The first passage is the functionalisation of 3D entities. To achieve this we have to characterize them in terms of how they behave. Then, to satisfy the second passage, we have to specify how something belonging to the fundamental ontology can evolve like a 3D entity. Section 3.2 will show how we can formalize this framework to deliver a precise account about the inter-level relation between the two realms. However, before making sense of the technical details of the functionalist scheme concerning the formal and metaphysical aspects, it is paramount to provide a clear picture of the account on the *physical* level. That is, our starting point will be to demonstrate via realistic cases how we can physically model wavefunctions as behaving three-dimensionally in the right circumstances, to highlight certain peculiar features of the functionalist approach to WFR, which will be crucial for the development of my functional reductionist account. Additionally, it should be said that if we cannot provide physical examples about the functional realization at stake then the functionalist approach would not be able to take off altogether, and thus it would be useless to discuss its formal and philosophical details. I will firstly consider a toy case – involving a single-particle quantum system[17] – and then move to a real world case, i.e. the Helium atom.[18,19] Finally, I discuss the case of decoherence.

---

[8] See Ney (2021b, pp. 39–40), Lewis (2004, pp. 715–16), and the introduction in Ney and Albert (2013).

[9] More accurately, not a configuration space *stricto sensu* (since now this is the fundamental space) but a space structurally isomorphic to it.

[10] Since the theory on which the present version of WFR is based on – i.e. non-relativistic quantum mechanics – is arguably superseded by better theories like quantum field theory, this should be read as a claim of *relative* fundamentality, or at least we should expect the thesis of WFR to be amended in order to accommodate quantum field theory. Keep this in mind whenever fundamentality is mentioned in the paper. See Ney (2021b) on the topic.

[11] In this sense, WFR qualifies as a form of *supersubstantivalism* (see Lehmkuhl (2018)). See Schroeren, 2022 (forthcoming) for a non-standard formulation of Core WFR eschewing supersubstantivalism.

[12] See e.g. Esfeld et al. (2014), Esfeld (2019), Lorenzetti (2021), and Chen (2019) for an overview.

[13] Actually from a quite specific version of it, using wavefunctions in position representation. Cf ft. 5.

[14] I stick here to the meaning of separability and locality adopted by Wave Function Realists (Ney, 2021b, ch. 3) and shared e.g. by Bell (2004).

[15] Notice: not *three*-dimensionally local, but local in the 3N-dimensional configuration space.

[16] On the contrary, the primitive ontology view cannot arguably save both features at the same time.

Take a quantum system associated with a wavefunction $\psi$, subject to a potential $V(x)$. We know the Schrödinger equation for the system:

$$i\hbar \frac{\partial \psi}{\partial t} = \hat{H}\psi, \tag{1}$$

where $\hat{H}$ is the self-adjoint Hamiltonian:

$$\hat{H} = \frac{\hat{p}^2}{2m} + V(\hat{x}) \tag{2}$$

Now, let's take a generic operator $\hat{Q}$, with associated expected value $\langle \hat{Q} \rangle$. We know, from Ehrenfest's theorem, that the time evolution of $\langle \hat{Q} \rangle$ can be stated as:

$$\frac{d}{dt}\langle \hat{Q} \rangle = \frac{i}{\hbar}\langle [\hat{H}, \hat{Q}] \rangle + \langle \frac{\partial \hat{Q}}{\partial t} \rangle \tag{3}$$

Now say that we have an isolated and localised wavepacket defined over configuration space. Its position, according to Ehrenfest's theorem, can be said to evolve in this way:

$$\frac{d}{dt}\langle \hat{x} \rangle = \frac{\langle \hat{p} \rangle}{m} \tag{4}$$

Similarly, for the momentum operator:

$$\frac{d}{dt}\langle \hat{p} \rangle = -\langle \frac{\partial V(\hat{x})}{\partial x} \rangle \tag{5}$$

Notice now that, on the assumption that $\langle \frac{\partial V(\hat{x})}{\partial x} \rangle$ is equal to $\frac{\partial V(\langle \hat{x} \rangle)}{\partial x}$, which is given by the fact that this is a localised wavepacket, the expectation values of the position and the momentum evolves like the classical position and momentum, and the last equation can be shown to be equal to Newton's second law. In fact, given what I have just said, we can write:

$$m\frac{d^2 \langle \hat{x} \rangle}{dt^2} = \frac{d\langle \hat{p} \rangle}{dt} = -\frac{\partial V(\langle \hat{x} \rangle)}{\partial x}, \tag{6}$$

which is, for narrowly localised wavepackets, equivalent to a high approximation to:

$$F = m\frac{d^2 x}{dt^2} = \frac{dp}{dt} = -\frac{dV(x)}{dx} \tag{7}$$

Notice that, within the quantum mechanical picture, the centre of the localised wavepacket has a trajectory is configuration space which is (to a very high approximation) identical to the trajectory in configuration space of a point particle of mass *m* within classical mechanics (in the Hamiltonian formulation). Thus, the trajectory of the wavepacket (as highlighted also by the Ehrenfest's theorem for the evolution of the momentum of the wavepacket) can be practically considered as a solution to the classical dynamic equation for a classical particle.

What does this mean? This means that a localised (isolated) wavepacket evolving in configuration space can behave – to a high approximation – as a classical point particle. More precisely, isomorphically to a classical point particle in configuration space; but for classical mechanics – differently from quantum mechanics (at least for the WFRist) – we have reasons to take that representation as simply a mathematical representation of a 3D particle in Euclidean space. And for the functionalist this is all we need. If all it takes to be a 3D particle is to behave according to Newton's law for the evolution of a point-particle, then we have just recovered a 3D particle from the evolution of a wavepacket.[20,21]

A flaw of this toy case is that this is a trivial case, since for *one* particle the configuration space is just 3D Euclidean space. However, as stressed also by Wallace (2012, p. 69), this case is easily extendible to a system of degrees of freedom to which we can apply the following Hamiltonian:

$$\hat{H} = \sum_{i=1}^{N} \frac{\hat{p}_i^2}{2m_i} + V(\hat{Q}_1, ..., \hat{Q}_N) \tag{8}$$

Then, as Wallace points out: "Localized wavepackets of this system will now pick out trajectories in a high-dimensional space, and these trajectories will instantiate the dynamics of a classical theory with N degrees of freedom" (p. 69).

Indeed, what I am going to do now is to introduce a case study for the functionalist account which is both non-trivial (since it concerns more than one particle, and thus configuration space is not the same as 3D Euclidean space), and also less abstract than the toy case I have just presented. In fact, I am going to apply the framework I have applied to the toy case to the real world case of the helium atom. Let's thus state the Hamiltonian of the helium atom as follows (dropping the hats):

$$H = \left\{-\frac{\hbar}{2m}\nabla_1^2 - \frac{1}{4\pi\varepsilon_0}\frac{2e^2}{r_1}\right\} + \left\{-\frac{\hbar}{2m}\nabla_2^2 - \frac{1}{4\pi\varepsilon_0}\frac{2e^2}{r_2}\right\} + \frac{1}{4\pi\varepsilon_0}\frac{e^2}{|\mathbf{r}_1 - \mathbf{r}_2|} \tag{9}$$

The first two terms of equation (9) are just like two hydrogenic Hamiltonians (with nuclear charge 2*e*), one for each electron in the helium atom, while the third term represents the repulsion between the two electrons. If we ignore the third term then we can rewrite (9) as follows:

$$H = \left\{-\frac{\hbar}{2m}\nabla_1^2 - \frac{1}{4\pi\varepsilon_0}\frac{2e^2}{r_1}\right\} + \left\{-\frac{\hbar}{2m}\nabla_2^2 - \frac{1}{4\pi\varepsilon_0}\frac{2e^2}{r_2}\right\}, \tag{10}$$

with potential energy [22]:

$$V(\mathbf{r}) = -\frac{1}{4\pi\varepsilon_0}\frac{2e^2}{r_1} - \frac{1}{4\pi\varepsilon_0}\frac{2e^2}{r_2}, \tag{11}$$

and we can then write the solution of Schrödinger's equation for the helium atom (in its ground state) as a wavefunction which is the product of two hydrogen wavefunctions[23]:

$$\psi_0(\mathbf{r}_1, \mathbf{r}_2) = \psi_{100}(\mathbf{r}_1)\psi_{100}(\mathbf{r}_2) = \frac{8}{\pi a^3} e^{-2(r_1 + r_2)/a} \tag{12}$$

---

[17] You can imagine it to be a Hydrogen atom, for instance.
[18] I assume here that the fundamental ontology is constituted just by the wavefunction, since I want to rely just on the most neutral textbook presentation of quantum mechanics. This is compatible with the Everettian and the GRW versions of WFR, but not with the Bohmian version. However, this is enough to sketch how the functionalist approach works.
[19] See Griffiths and Schroeter (2018, ch. 3, 5.2) and Sakurai and Commins (1995, p. 86).
[20] Within classical mechanics every quantity is fixed by the position and the momentum, so here we have really obtained a full-fledged classical particle. Notice that the point is just to get classical mechanics from non-relativistic quantum mechanics.
[21] In this sense, the identification between the expectation values for e.g. position and the position variable is justified by functionalism, i.e. by the claim that if the variable $\langle \hat{x} \rangle$ in the models of quantum mechanics dynamically evolves (to a high approximation, for localised wavepackes) like *x*, then the two variables can be identified in that context.
[22] Notice that this is just the classic Coulomb's potential.
[23] Actually, if we simply ignore the repulsion between the electrons, the energy we can calculate from Hamiltonian (10) will not correspond to the experimentally measured energy, since the repulsion is not negligible. However, there are no analytic solutions to Schrödinger's equation for the Helium atom. We can get a better result if we employ approximation methods and we use, instead of (12), a trial wavefunction such as: $\psi_0(\mathbf{r}_1, \mathbf{r}_2) = \frac{Z^3}{\pi a^3}e^{-Z(r_1+r_2)/a}$ (where Z is a variational parameter that we get from the approximation procedure) to which we can assign the following Hamiltonian, instead of (9): $H = -\frac{\hbar}{2m}(\nabla_1^2 + \nabla_2^2) - \frac{e^2}{4\pi\varepsilon_0}(\frac{Z}{r_1} + \frac{Z}{r_2}) + \frac{e^2}{4\pi\varepsilon_0}(\frac{Z-2}{r_1} + \frac{Z-2}{r_2} + \frac{1}{|\mathbf{r}_1-\mathbf{r}_2|})$. However, this is not crucial for the purpose of the present discussion, so I shall stick with equations (10–12).

Notice that this wavefunction depends on both $\mathbf{r}_1$ and $\mathbf{r}_2$ and thus is defined over a 6-dimensional configuration space. Thus, it is not the same as the previous case for the single particle, where the space on which the wavefunction was defined was 3-dimensional just like Euclidean space. The crucial point now is the application of Ehrenfest's theorem to this quantum system. The theorem allows us to say that the expectation value for the position operator of the first electron $\langle \mathbf{r}_1 \rangle$ (and the same can be said for the second one) evolves classically as $\frac{d\langle \mathbf{r}_1 \rangle}{dt} = \frac{\langle \mathbf{p}_1 \rangle}{m}$, and consequently the evolution of the momentum is:

$$\frac{d\langle \mathbf{p}_1 \rangle}{dt} = -\nabla V(\langle \mathbf{r}_1 \rangle), \tag{13}$$

provided that the wavefunction of electron is sufficiently localised in space. Thus – when the conditions we have set are satisfied – the electrons in the helium atom (approximately) behave according to Newton's equation for the evolution of a particle subject to an electric potential. If the condition to be a classical particle is to evolve according to Newton's equation, then we can say that – in the limit case – the 6-dimensional wavefunction of the two electrons in the helium atom turns out to correspond to two classical 3D point particles.

Having presented the main case study, a consideration concerning the limits of this example is in order. The example discussed so far rests on a few strong assumptions, as we considered the behaviour of localised and isolated wavepackets. These premises are not simply guaranteed to obtain in most realistic contexts, and in general they may seem to require a too high degree of idealisation. In particular, quantum systems are not usually isolated, and interference phenomena can pose a threat to the recovery of classicality too. What solves the problem is the appeal to decoherence.[24] I briefly present here – following Zurek (2006).[25] – how decoherence guarantees the suppression of the interference and the emergence of quasi-classical behaviour in the case of non-isolated systems of a very general kind. Take a system in a pure state $|\psi_s\rangle = \alpha|\uparrow\rangle + \beta|\downarrow\rangle$, with $|\alpha|^2 + |\beta|^2 = 1$, and consider also a quantum detector whose Hilbert space is spanned by states $|d_\uparrow\rangle$ and $|d_\downarrow\rangle$. The detector is built in such a way that it starts in state $|d_\downarrow\rangle$ and is transformed in the following way $|\uparrow\rangle|d_\downarrow\rangle \to |\uparrow\rangle|d_\uparrow\rangle$ when the measured state is $|\uparrow\rangle$ and remains unchanged otherwise. The composite system before the interaction is $|\varphi^i\rangle = |\psi_s\rangle|d_\downarrow\rangle$. Interaction causes $|\varphi^i\rangle$ to transform into $|\varphi^c\rangle$ as follows:

$$|\varphi^i\rangle = (\alpha|\uparrow\rangle + \beta|\downarrow\rangle)|d_\downarrow\rangle \to \alpha|\uparrow\rangle|d_\uparrow\rangle + \beta|\downarrow\rangle|d_\downarrow\rangle = |\varphi^c\rangle \tag{14}$$

In this situation, the density matrix of the pure state would be:

$$\rho^c = |\varphi^c\rangle\langle\varphi^c| = |\alpha|^2|\uparrow\rangle\langle\uparrow||d_\uparrow\rangle\langle d_\uparrow| + \alpha\beta^*|\uparrow\rangle\langle\downarrow||d_\uparrow\rangle\langle d_\downarrow| + \alpha^*\beta|\downarrow\rangle\langle\uparrow||d_\downarrow\rangle\langle d_\uparrow| + |\beta|^2|\downarrow\rangle\langle\downarrow||d_\downarrow\rangle\langle d_\downarrow|, \tag{15}$$

where the two middle (off-diagonal) terms represent the interference. Recovering classicality would require the off-diagonal terms to be cancelled out, transforming $\rho^c$ into[26]:

$$\rho^r = |\alpha|^2|\uparrow\rangle\langle\uparrow||d_\uparrow\rangle\langle d_\uparrow| + |\beta|^2|\downarrow\rangle\langle\downarrow||d_\downarrow\rangle\langle d_\downarrow| \tag{16}$$

The point is that $\rho^r$, contrarily to $\rho^c$, simply represents 'classical ignorance' about the state. The reduced density matrix can be said to represent alternative classical states of the detector system which we are just ignorant about. Decoherence shows how this can take place, by looking at the interaction between the state and the environment. Let's call the system, the detector and the environment respectively $\mathcal{S}, \mathcal{D}, \mathcal{E}$.

The system $|\varphi^c\rangle$, that includes the detector, can be said to interact with the environment $\mathcal{E}$ as follows:

$$|\varphi^c\rangle|\mathcal{E}_0\rangle = (\alpha|\uparrow\rangle|d_\uparrow\rangle + \beta|\downarrow\rangle|d_\downarrow\rangle)|\mathcal{E}_0\rangle \to \alpha|\uparrow\rangle|d_\uparrow\rangle|\mathcal{E}_\uparrow\rangle + \beta|\downarrow\rangle|d_\downarrow\rangle|\mathcal{E}_\downarrow\rangle = |\psi\rangle \tag{17}$$

The final state extends the quantum correlation from the pair $\mathcal{SD}$ to $\mathcal{SDE}$. Now, as Zurek (p. 39) stresses, if the states of the environment $|\mathcal{E}_i\rangle$ corresponding to states $|d_\uparrow\rangle$ and $|d_\downarrow\rangle$ of the detector are orthogonal,[27] we can trace over the degrees of freedom corresponding to the environment such that we can obtain the following density matrix describing the $\mathcal{SD}$ combination:

$$\rho_{\mathcal{SD}} = Tr|\psi\rangle\langle\psi| = \sum_i \langle \mathcal{E}_i|\psi\rangle\langle\psi|\mathcal{E}_i\rangle \cong \rho^r \tag{18}$$

We have thus (approximately) obtained $\rho^r$ as needed. The correlation has now spread over to the environment. However, if we focus on our systems of interest – since we can neglect the environment's degrees of freedom, for all practical purposes – we can claim that the components of the system-plus-detector state have become effectively dynamically decoupled. Cutting off the off-diagonal elements in the density matrix of the system entails that the system has effectively lost coherence as interference has been effectively canceled out, and the wave function corresponds to distinct localized and internally stable wavepackets, to a sufficient degree of approximation.[28] We have recovered quasi-classicality in a crucial sense, and in fact $\rho^r$ just represents ignorance about possible classical states.[29] Without peering further into the topic, let's stress how this phenomena is connected with our discussion.[30] As Wallace (2010) points out, if we denote with $|q,p\rangle$ a state of a system localised around $(q, p)$ in phase space, decoherence then ensures that we can write the system + environment state at any time $t$ as:

$$|\psi\rangle = \int dq dp \, \alpha q,p;t|q,p\rangle \otimes |\varepsilon(q,p)\rangle \tag{19}$$

with $\langle \varepsilon(q,p)|\varepsilon(q',p')\rangle = 0$, unless $q \approx q'$ and $p \approx p'$. Textbook quantum mechanics says that $|\alpha(q,p)|^2$ is the probability density for finding the system near $(q, p)$. Crucially, in presence of decoherence, $|\alpha(q,p)|^2$ evolves to a high approximation as a classical probability density on phase space, i.e. under Poisson equations:

$$\frac{d}{dt}(|\alpha(q,p)|^2) \simeq \frac{\partial H}{\partial q}\frac{\partial |\alpha(q,p)|^2}{\partial p} - \frac{\partial H}{\partial p}\frac{\partial |\alpha(q,p)|^2}{\partial q}, \tag{20}$$

with Hamiltonian $H(q, p)$ – as happens in the case of localised wavepackets in the context of Ehrenfest's theorem discussed before. Indeed, Wallace highlights that:

> On the assumption that the system is classically non-chaotic (chaotic systems add a few subtleties), this is equivalent to the claim that each individual wavepacket follows a classical trajectory on phase space. Structurally speaking, the dynamical behaviour of each wavepacket is the same as the behaviour of a macroscopic classical system. And if there are multiple wavepackets, the system is dynamically isomorphic to a collection of independent classical systems (Wallace, 2010, p. 64).

---

[24] Thanks to an anonymous reviewer for suggesting to address the topic. See Wallace (2012, pp. 74–76) for more details on the limits of Ehrenfest's theorem.
[25] See also Zurek (2003) for more details.
[26] This is the result we would get from adding a collapse postulate to the dynamics of quantum mechanics, i.e. collapse would eliminate one term of the superposition and turn $\rho^c$ into a proper mixture.
[27] See Schlosshauer (2005, p. 10) on the justification of the orthogonality premise.
[28] As remarked by Thébault and Dawid (2015): "Environment induced decoherence does not fully eliminate the off diagonal elements, but it re-scales them to vanishingly small amplitudes as given by the associated Born weights" (p. 1565). I will come back to the topic of approximation in the following.
[29] See on this Wallace (2010, p. 61).
[30] A complete analysis of decoherence would require a discussion about how different quantum theories incorporate this phenomenon (see Schlosshauer (2005)). Since I want to keep this presentation as neutral as possible with respect to versions of quantum theory, I just stick to the decoherence mechanics common to every approach, as this is already key to show how we can recover classicality from quantum states.

Thus, we can recover the dynamical conditions we are looking for even in the – less idealised – cases of non-isolated interacting systems.

To conclude our presentation, I stress three considerations about the assumptions used throughout this subsection. First, these cases require a certain amount of idealisation and approximation. However, my aim here is just to show in a technically perspicuous way how the functionalisation works – thus, this is enough to give you a first idea of the mathematical details behind the functionalist idea. This is already a decisive step forward concerning the formulation of functionalism within the WFR debate. It is because of this that I shall focus more on the Ehrenfest-type situations, as they are more tractable and suit our aims nonetheless. Second, relatedly, the appeal to approximation is actually a *virtue* of the functionalist account. Approximation is used all over the place in physics, and the fact that functionalism just requires approximation of behaviour (up to what it matters at the practical level) is a virtue of functionalism over alternative reductive modes of recovering the 3D realm which may require exact correspondence or exact reduction. I discuss further the combination between functionalism and approximate reduction in section 3.3. Third, and crucially, WFR already requires highly localised (peaked) wavefunctions in configuration space to have the 'emergence' of the 3D ontology. So the use of Ehrenfest or decoherence is a natural fit within the functionalist framework proposed by wavefunction realists, and, in any case, is based on an assumption which is already shared also by those defenders of WFR who don't support functionalism (e.g. Ney (2021b)).

To recap, the functionalist models I have just presented give us reasons to believe that a functionalist approach to WFR is viable, and they also show us the way in which the functionalisation works within WFR. This approach relies on semiclassical features and, in the way I have described it, it is just based on standard textbook features of quantum mechanics.

## 3.2. Lewisian functional reduction

Section 1.1 introduced Minimal Functionalism, while last section showed how some actual physical examples can fit in that framework. The next step towards the introduction of my functional reductionist framework is then to provide a more formalised version of Minimal Functionalism. Drawing on the account defended by Lewis (1970, 1972) concerning theoretical terms and functionalism, and on some plausible assumptions, I will show that a form of functional reductionism – that bears important consequences to the whole debate about WFR – follows from Minimal Functionalism about WFR.

The main role of the Lewisian framework here is that of giving us a formal way to functionally characterize the specific entities that we want to functionalise. Indeed, our task can be, e.g., to provide a functionalisation of a two-particle classical system like the one we saw in the context of Ehrenfest's theorem. I will take that as my main example later in the section, to show in more details how the account works. The reason is that this is a fairly tractable case study which we have also discussed in details at the physics level. A discussion about decoherence contexts and how they fit within the functionalist picture will follow the main case study.

Let's proceed carefully, and start from the general Lewisian functionalist framework. This account can be used to build a functional definition for a given (or every) theoretical entity which is embedded in a theory. Take thus a physical theory and call *T*-terms the theoretical terms $\tau_1, \ldots \tau_n$ introduced by the theory, and call the rest of the terms in which the theory is couched *O*-terms. Let's then form the *postulate* of the theory *T*:

$T(\tau_1, \ldots \tau_n)$

This is a sentence that contains all the theoretical postulates of the theory (e.g. $\vec{F} = m\vec{a}$), expressed as a long conjunction. If we replace the *T*-terms with open variables, we obtain the *realization formula* of *T*:

$T(x_1, \ldots x_n)$

Any *n*-tuple of entities that satisfies this formula may be said to realize the theory *T*. We can now introduce the *Ramsey sentence*, which says that *T* is realized:

$\exists x_1, \ldots x_n T(x_1, \ldots x_n)$

Then, assuming that the Ramsey sentence is uniquely realized we can build functional definition for the theoretical terms:

$\tau_1 \stackrel{def}{=} \iota y_1 \exists y_2, \ldots y_n \forall x_1, \ldots x_n (T[x_1, \ldots x_n] \equiv . y_1 = x_1 \wedge \ldots \wedge y_n = x_n)$

$\tau_n \stackrel{def}{=} \iota y_n \exists y_1, \ldots y_{n-1} \forall x_1, \ldots x_n (T[x_1, \ldots x_n] \equiv . y_1 = x_1 \wedge \ldots \wedge y_n = x_n)$

Assuming that our theory is uniquely realized, i.e. that there is just one set of entities that realize the theory, then once we write down the postulate of the theory and we derive the realization formula, we can derive an explicit definition for each of the theoretical terms in the theory. As Lewis claims:

> This is what I have called functional definition. The *T*-terms have been defined as the occupants of the causal roles specified by the theory *T*; as *the* entities, whatever those may be, that bear certain causal relations to one another and to the referents of the *O*-terms (Lewis, 1972, p. 255).

Thus, this is a formal way functionally characterize the entities postulated by a certain theory. However, notice that this kind of account can be used also to functionally characterize just one or few entities, without having to replace every theoretical term in *T* with a variable. Indeed, let's say for instance that there is just one 'problematic' theoretical term that we want to functionally define (e.g. 'point particle' or 'two-particle system'). Then, we can replace just that term with a variable and construct in this way the Ramsey sentence. At that point, we can have just one functional definition of the form '$\tau \stackrel{def}{=}$'. That definition would say that that precise term $\tau$ is just whatever thing that fills that causal role within the theory, i.e. that thing that bears those relations with the other (un-functionalized) terms of the theory.[31]

Let's set this aside for a moment, and suppose now that a second theory *T** is introduced. This will be our 'bottom' theory. *T** introduces a new set of theoretical terms, which we can call *O**-terms. *O**-terms are either *T**-terms or *O*-terms. Suppose further that:

$T^* \vdash T[\rho_1 \ldots \rho_n]$

where $\rho_1 \ldots \rho_n$ are *O**-terms, introduced independently from the terms $\tau_1, \ldots \tau_n$. $T[\rho_1 \ldots \rho_n]$ is called the *weak reduction premise* for *T*, and it does not contain *T*-terms. It says that *T* is realized by a *n*-tuple of entities $\rho_1 \ldots \rho_n$. Thus, *T* is realized by a *n*-tuple of entities expressed in the vocabulary of the new theory. Now, Lewis points out that the postulate $T(\tau_1, \ldots \tau_n)$ can be derived from the weak reduction premise together with some bridge laws of the following form, which are usually taken as separate empirical hypotheses:

$\rho_1 = \tau_1, \ldots, \rho_n = \tau_n$

Alternatively, the bridge laws can be derived from *T** alone. In the case in which *T* is uniquely realized by a *n*-tuple of entities named by $\rho_1 \ldots \rho_n$, we can accept the following sentence, which we can call the *strong reduction premise* for *T*:

---

[31] Notice that if we adopt this strategy, and we functionalise just one or just some specific terms within the theory, and not the whole theory at one, we are able to contrast the standard Newman's objection that could be raised against the Ramsey sentence method (see e.g. Demopoulos and Friedman (1985)).

$$\forall x_1, \ldots x_n (T[x_1, \ldots x_n] \equiv . \rho_1 = x_1 \wedge \ldots \wedge \rho_n = x_n)$$

This sentence logically implies the following definitions, which are $O^*$-sentences and can be therefore theorems of $T^*$:

$$\rho_1 = \iota y_1 \exists y_2, \ldots y_n \forall x_1, \ldots x_n (T[x_1, \ldots x_n] \equiv . y_1 = x_1 \wedge \ldots \wedge y_n = x_n)$$
$$\rho_n = \iota y_n \exists y_1, \ldots y_{n-1} \forall x_1, \ldots x_n (T[x_1, \ldots x_n] \equiv . y_1 = x_1 \wedge \ldots \wedge y_n = x_n)$$

Which entails the theoretical identifications $\rho_1 = \tau_1, \ldots, \rho_n = \tau_n$ by transitivity of identity. That is, the strong reduction premise entails the theoretical identifications by itself. The crucial point to highlight here is that the theoretical identifications between the two theories follows *deductively*. That is, if we adopt a functionalist understanding of theories, once we functionalise a theory and then we find out another theory (possibly with even more explanatory power), whose entities can realize the former theory, we are committed to draw theoretical identifications across the two theories. If an entity belonging to the bottom theory plays the role that we associated to another entity within the top theory, those two entities are identical (entailing a form of 'realizer' functionalism).

Now, notice that I expressed this second passage about reduction in terms of a whole set of theoretical identifications. However, as I mentioned a few paragraphs above, concerning the construction of the functional definitions within the context of the top theory, we could apply the functionalisation strategy just to one single theoretical entity – or a few of them. However, this does not affect the second passage: if that is the case, in the moment in which we are writing down e.g. the strong reduction premise for $T$, we will have just one identification, based on which exact element in the bottom theory plays that specific role played by $x$, and thus we will end up with just one theoretical identification eventually.

Let's now apply the Lewisian account to WFR, to show how it can formalize the Minimal Functionalist approach. In particular, I mostly focus on the functionalisation of the kind of entity we have studied in our example related to Ehrenfest's theorem in section 3.1. That is, I show how we can functionalise the classical notion of 'two-particle 3D system'.[32] This allows us to study how the functionalist account can be spelled out in a realistic and detailed case. A more general discussion about more general cases and decoherence within WFR will follow.

First, take classical mechanics as our theory of the 3D world,[33] and take that as our top theory. Now, it seems sensible to hold that in our universe classical mechanics is uniquely realized, assuming it is approximately true, at least in a certain domain. No two sets of entities in our universe can really realize the theory. Thus, we can use the first step of the Lewisian account to build functional definitions for classical mechanics. In particular, we can use it for example to define functionally the notion of 'two-particle 3D system'. In other words, this is a formal way to spell out statement (i) within the Minimal Functionalist thesis, and to clarify the notion of 'functional understanding' of 3D objects.[34] In our case, the very notion of 'two-particle 3D system' – i.e. what it takes for something to be a two particle 3D system – is completely spelled out in functionalist terms, in the sense that it is completely (functionally) defined in a precise and unambiguous way in terms of its role within the theory, via the Ramsey sentence. Having functionalized this 3D entity (the classical system) we can move to the second step of the Lewisian account, i.e. we can consider our bottom theory, that is quantum mechanics. I have shown before how one can hold that the wavefunction – evolving in its space – can behave in an approximatively 3D way, thereby playing the role of 3D entities. We have seen this in particular in the regime of Ehrenfest's theorem, where our specific kind of 6-dimensional wavefunction approximatively evolved according to Newton's equation in a particular regime. Thus, going back to the general picture, if we take non-relativistic quantum mechanics to be the theory $T^*$, and we grant that quantum mechanics is uniquely realizable, and we also grant that $T^*$ specifies an ontology which can indeed play the role which was played by 3D particles in classical mechanics, then following the Lewisian functionalist account we are forced to draw (type) identities between elements of the high-dimensional ontology of quantum mechanics and the 3D objects. Turning again to our case study, we are now in the position to say that functionalism, as applied to our two-particle 3D system, eventually entails an identity between 'two-particle 3D system' and 'highly localised 6-dimensional wavepacket'.

To be more precise about the relata of this relation, let's look back at the way in which I characterised the fundamental ontology of WFR. In section 2 I presented the view as committed to the claim that what exists fundamentally are high-dimensional points of the configuration space instantiating certain properties, as represented in the wavefunction. Following the functional reductionist account presented here, in our case we can say that a two-particle system turns out to be identical with a single point in the high-dimensional configuration space.[35] We could say that the three-dimensional system is, after all, a redescription of the fundamental high-dimensional entity. This is an admittedly radical conclusion, but it follows deductively from the functionalist account, once we show (as we did) that the two sorts of entities – i.e. the system and the configuration space point – fulfill exactly the same causal role. If one wants to avoid this conclusion, one has to either discard Minimal Functionalism, or prove where the argument goes wrong – either at the level of mathematical details or concerning the Lewisian formalization of Minimal Functionalism. In any case, I believe that the discussion carried out from section 3.1 to this point already serves the important purpose of making clear the details of the functionalist account of WFR in both the mathematical and the philosophical aspects, by considering this account with respect to a specific case study, thereby filling a gap in the literature. In this sense, the account presented here does not *transform* functionalism about WFR into a radical view, but rather it shows the radical consequences of the view once one is precise about the details. However, section 4 and 5 will show that this functional reductionist account and the identity relations it entails lead also to positive features.

To wrap up, this section showed via our case study that the metaphysical relation between the two levels is actually *identity*. This important fact has never been pointed out in the literature on WFR,[36] but it is a very natural conclusion once we maintain both that 3D entities should be functionally defined and that the wavefunction functionally enacts them. Actually, this conclusion follows deductively, once one adopts Minimal Functionalism about WFR, as we have shown thanks to the Lewisian formalization.

At this point, as we did in section 3.1, it should be acknowledged that the case which we have focused on to show the details of functional reductionism within WFR is rather limited. As I stressed, a realistic treatment of the dynamics of quantum states would require an appeal to decoherence phenomena. We saw in the previous subsection the physics behind decoherence, and I pointed out how this mechanics make sure that approximately classical behaviour comes into place in the case of interacting systems. How can we embed that within the Lewisian framework? At the general level, to satisfy the account presented here and provide functional reduction, we just need to show how a quantum system can behave like a classical one in the right regime. Looking at section 3.1, the description of decoherence presented there provides what we need. As I showed, after the interaction with the environment, interference effects in the quantum state are effectively eliminated and we can trace out the environment's degrees of freedom for all practical purposes, and we are thus able to demonstrate that the quantum system under analysis can then behave according to classical equations, like we did for the two-particles case. If we are interested in e.g. a single-particle system interacting with the environment, decoherence ensures that the system will eventually behave classically, i.e. follow a classical trajectory in phase space. What we cannot however plausibly do in *every* situation, is to draw an exact correspondence between higher-dimensional and lower-dimensional degrees of freedom in forms of identity relations. Decoherence is often predicated at macroscopic and coarser levels of description, where the quantum systems are complex and the degrees of freedom are countless. If the quantum system we are focusing on is a complex entity, it is going to be practically very challenging to track all of its degrees of freedom, and thus trying to draw identity relations of the kind we specified in our case study would be a hopeless task. Furthermore, we have seen that in the context of decoherence superposition does not really go away, but is spread out, and thus one may ask whether this can block the cross-level identifications.[37] Nevertheless, I argue that these concerns should not trouble us.

---

[32] In particular, a two-particle 3D system is an idealised system composed by two classical point-particles. The idea is to show how the wavefunction of e.g. a helium atom can eventually enact two ideal point-particles that evolve under classical equations like two point-particles in the theory of classical mechanics would do. As mentioned in the previous subsection, the appeal to idealisation is motivated by the need to deal with a manageable model, to provide a simple example of the functionalist account.

[33] This is not exactly correct, but take this for granted for a moment – see section 3.3 for more details.

[34] One could say that the concept of 'being 3D' has features which are not related to non-relativistic classical mechanics. However, I take here 'being 3D' as a theoretical concept. Given that playing the role associated with the theoretical concept of three-dimensionality is plausibly at least a *sufficient* condition to be 3D in general, then this is all we need for our purpose, i.e. to recover 3D entities in the last place.

[35] Other possible non-supersubstantivalist readings of the fundamental ontology of WFR will disagree on the exact relata, however the point of this passage is just to show in more details which kind of consequences does the functional reductionist approach lead to.

[36] Although it has been recently explicitly advocated by Huggett & Wüthrich, (2021) concerning functionalism as applied to spacetime in quantum gravity. Cf. also Huggett and Wüthrich (2013).

First of all, the account does not require us to provide precise mappings between the exact degrees of freedom of the high-dimensional and the low-dimensional spaces. All we essentially need is to show how the system denoted by the top theory (classical mechanics) can be functionally realized by the quantum system. The previous subsection showed how this works physically, and thus this would be enough to satisfy our desiderata. Second, a reason to rest satisfied with this more coarse-grained kind of identity relation is given by the fact that the concept of macroscopic system might even be too vague to allow for an exact precisification of its composition.[38] But that would be more of a pragmatic or linguistic problem than a real metaphysical issue of the kind that should trouble us. Allowing for a coarser kind of identity relation seems reasonable when we deal with less precise concepts. Another aspect of this consideration is that, in general, we don't expect to be able to describe complex quantum systems via precise wavefunctions. At best, we can provide approximate descriptions, which would not allow us to track down the exact distribution of the wavefunction over configuration space, for pragmatic reasons. However, the fact that we were able to do so in the simple case of the two-particles isolated system strongly suggests that we would be able to do so also for more complex systems if we had unlimited knowledge about the world: given that decoherence – similarly to Ehrenfest's theorem – provides us with the right dynamics, complex systems are in principle able to meet the criteria for classicality fixed by functional reductionism, and thus simple and complex system just differ on an epistemological point of view. Finally, a point about the interaction between systems and the environment. When dealing with the simple isolated system, it was easy to track the correspondence between the quantum and classical description of the system in the regime of interest, and the main approximation we relied on concerned the trajectory of the localised quantum system in phase space, which was said to approximate a classical trajectory. If we move to more complex contexts, like the ones in which decoherence plays a role, we are naturally going to require more assumptions and more approximations. One of them concerns the fact of tracing out and neglecting the environment. That approximation is validated by scientific practice, and thanks to it we are justified to ignore the environment's degrees of freedom and the lingering superposition relations. Because of this, we can reasonably just focus on the decohered system when discussing the functional realization relation, and it's just the system that is going to turn out as identical to a classical one in that context.[39] All in all, when we move from simple limited examples to complex contexts where decoherence is needed, the complexity increases, and the precision we can achieve in drawing our identity relations decreases, but nothing in this process poses substantive challenges to the Lewisian account.

The next subsection elaborates on some further important details of the functional reductionist account introduced here, to clarify the position and to anticipate possible objections. In particular, I expand on the role of approximation within the theory, and on the way in which the identity relations are brought about within the account, focusing on our case study.

### 3.3. Further remarks: approximation and identity

This subsection ties together the loose ends of the last ones and elaborates on the functional reductionist account of WFR. I first clarify a point concerning the reductionist aspect of the Lewisian functionalist framework with respect to approximation, and then add some remarks about the nature of the identity relations discussed in the last section.

First of all, notice that I have taken classical mechanics as our top theory *T* within the Lewisian schema. However, this seems to be in clear contrast with the physical details discussed in section 3.1. For instance, in the case of Ehrenfest's theorem, the highly localised wavepackets do not behave exactly according to Newton's equation, but rather obey an approximate version of it, and something similar happens with decoherence. For this reason, classical mechanics seems unsuitable as top theory. However, a quite straightforward solution can be borrowed from the literature about reductionism. In fact, following the so called Neo-Nagelian brands of reductionism (cf. Schaffner (1967) and Dizadji-Bahmani et al. (2010)), we can claim that, to ensure reduction, we actually just need to replace our original top theory with another theory standing in a relation of good approximation with the former.[40] In our case, this approximate version of classical mechanics, that contains only an approximate version of Newton's equation, and that we can use to functionally define the 3D entities, is now perfectly suitable for our purpose. In other words, replacing the exact version of classical mechanics with an approximate version of the theory accommodates the idea, expressed throughout the last sections, that the functional reduction of the 3D ontology to the high-dimensional one is to be expressed in terms of wavefunctions behaving *approximatively* classically.

The second point to highlight concerns the way in which the identity relations between high-dimensional and 3D entities are produced. Notice that identities are established only in the appropriate limited contexts in which can practically take the wavefunction as 3D. In fact, according to this account, the identity between states of the wavefunction and 3D entities does not hold always and in every situation: they hold only in those situations in which the wavefunction plays the right roles. That is, when the wavepackets evolve according to (approximate) classical laws, like it happens in the application domain of Ehrenfest's theorem, or when decoherence cancels out interference. In all the other cases the identity conditions are not satisfied and we don't have any 3D object. This is neither troubling nor unexpected: what exists fundamentally is just the high-dimensional ontology, after all. The conditions for identity are thus special cases in which we can redescribe the fundamental ontology three-dimensionally.

---

[37] Thanks to an anonymous reviewer for pointing out these issues to me.
[38] By talking of 'coarse-grained identities' I am not postulating a new special kind of relation, but rather I am just referring to the fact that we are identifying the systems at the more general macro-level of whole systems instead of focusing on identifying the specific degrees of freedom, as mentioned above.

[39] Besides this, recall also that decoherence should be implemented with a solution to the measurement problem anyway, and that would definitely help to tackle superposition.
[40] Notice that this approach is orthogonal to the debate about the semantic and the syntactic view of theories, and thus the approximation could easily be a relation between models.

The same consequence follows from the application of Lewisian functionalism to the philosophy of mind. That is, neural states are identical to mental states only when they are playing the appropriate roles. There is an asymmetry between the two ontologies. There can be states of the brain which do not correspond to any mental state, but not vice versa. If there is a mental state, it is because we have a specific neural state which is playing the right role which is need to bring about that mental state. In other words – in a relative sense of fundamentality – only brain states exist fundamentally, while mental states come into existence only in specific occasions in which the brain acts in certain ways. Only when the brain acts in such a way to play the role of pain, then the pain mental state exists, and its identical to the corresponding neural state.

Why is this important? First, this consideration will be useful to address an objection raised by Ney that will be discussed in section 5. Second, it is crucial to anticipate a possible critique that could be moved to the version of WFR proposed here. Recall that in section 2 I have said that a central advantage of WFR, according to its defenders, is the ability of the theory to maintain both locality and separability at the fundamental level. Then, given a reductionist account which entails identity relations between high-dimensional and 3D entities, one could complain that this version of WFR has to discard fundamental locality and separability, because there would be situations in which the wavefunction happens to be identical to 3D entities – e.g. two 3D particles in a singlet state – which in turn are not supposed to be separable and can also violate locality, and in those cases also the high-dimensional ontology would turn out to be non-local and non-separable, due to the relation of identity.

However, if we pay attention to how identities are brought about, we see that the problem does not arise in the first place. The reason is that the very conditions that we have set for being a 3D system based on classical behaviour rule out the possibility of non-separable non-local kinds of 3D systems. As I stressed, we have identity relations only in those specific situations in which the wavefunction behaves approximately classically, and that is related to being highly localised in the quantum phase space in a way such that the system can approximately evolve according to a classical trajectory. Thus, when a wavefunction is spread in configuration space and is not highly localised, then that wavefunction will not correspond to any 3D entity: in those situations all we have is the wavefunction in its high-dimensional space. That entity is separable and local for the reasons given by WFR, as presented in section 2. On the contrary, 3D systems come into play only in those situations in which the wavefunction is sufficiently highly localised, and in those cases – for all practical purposes, and to a high approximation – the 3D system will be both separable and local. That is, a measure on it would not violate locality, and separability stands too. In other words, we have fundamental separability and locality in all the situations in which we need separability and locality to hold. Thus, the functional reductionist version of WFR enjoys the same advantages of the general version of the theory.

This concludes the presentation of the functional reductionist account of WFR. In the next two sections I will first introduce Albert's Primitive Ontology Functionalism and then argue that my account avoids several objections that can be raised against his account and against any functionalist approach to WFR in general.

## 4. Albert's Primitive Ontology Functionalism

Having introduced my functionalist proposal, it is now time to point out that also Albert (2015) has proposed a functionalist account of WFR going beyond Minimal Functionalism. Differently from my account, which is neutral with respect to the solution to the measurement problem one can adopt and just relied on features of orthodox textbook quantum mechanics, Albert proposes a slightly different version of WFR for each version of quantum mechanics. In what follows I will present the functionalist account he proposes for GRW theory (but keep in mind that the general structure of the proposal can be equally applied to e.g. Bohmian mechanics), and then argue why the account I propose is preferable.

According to Albert, in the GRW version of WFR we can set out a wavefunction in the $3N$-dimensional space and an emergent (e.g.) matter-density field in the 3D space. To explain how the wavefunction can functionally bring about the low-dimensional ontology, we should go as follows. First, consider the fundamental $3N$-dimensional space and call it $S$, and then suppose that we can define a coordinatization $C$ of this space, which individuates a 3D subspace of $S$ as spanned by the $C$-axes $(3i - 2, 3i - 1, 3i)$, for $i = 1, 2, \ldots, N$. Then consider a function $f_i(x_{3i-2}, x_{3i-1}, x_{3i})$ which takes as input positions in the 3D subspaces and gives as output the amplitude of the wavefunction at that positions. Call the function $f_i(x_{3i-2}, x_{3i-1}, x_{3i})$ the $i$-th *shadow* of the wavefunction. For each collapse of the wavefunction there is a shadow in the 3D subspace. The challenge is to prove that shadows can behave as 3D particles. To do so, Albert first introduces an Hamiltonian relative to the coordinatization $C$:

$$H = \sum_i m_i \left( \left( \frac{dx_{3i-2}}{dt} \right)^2 + \left( \frac{dx_{3i-1}}{dt} \right)^2 + \left( \frac{dx_{3i}}{dt} \right)^2 \right) + \sum_i \sum_j V_{ij}((x_{(3i-2)} - x_{(3j-2)})^2 + (x_{(3i-1)} - x_{(3j-1)})^2 + (x_{(3i)} - x_{(3j)})^2), \quad (21)$$

where $i$ and $j(i, j = 1, 2, \ldots, N)$ range over the shadows. Then he defines the 3D Hamiltonian according to which the mass-density 3D field of GRW should evolve:

$$H = \sum_i m_i \left( \left( \frac{dx_i}{dt} \right)^2 + \left( \frac{dy_i}{dt} \right)^2 + \left( \frac{dz_i}{dt} \right)^2 \right) + \sum_i \sum_j V_{ij}((x_i - x_j)^2 + (y_i - y_j)^2 + (z_i - z_j)^2). \quad (22)$$

Then, his strategy is basically to confront (21) and (22) and pointing out the similarity in their forms: if the wavefunction evolves accordingly to the Hamiltonian (21), then the coordinates of the subspaces where the shadows live $(3i - 2, 3i - 1, 3i)$ will play the role of the coordinates $x, y, z$ in the Hamiltonian (22) for the 3D elements of the ontology. Thus, the shadows can be said to behave like 3D objects, and then GRW theory "is going to accommodate relatively stable 3D coagulations of subsets of these shadows in the shapes of tables and chairs" (pp. 137–138), and these shadows will play the role of tables and chair, and thus they will *be* tables and chairs. This account gives us, in a package, an alleged explanation of how we can physically take the wavefunction to play the role of 3D objects *and* a commitment about the nature of the 3D ontology and its relation with the wavefunction. In fact, the 3D ontology at stake is just the same as the one postulated by primitive ontology view for GRW,[41] with the caveat that in Albert's account it is not fundamental, and the relation between the two is a form of 'emergence'. I call this package *Primitive Ontology Functionalism*.[42]

However, some doubts have been cast on his proposal. I focus here on the issues related to the physical model, while I will discuss the

---

[41] I have presented here a version of it in terms of emergent matter-density field, but the same account can be formulated in terms of a 'flash' ontology (Albert (2015), ch. 6).

[42] Notice that this approach is importantly different from Esfeld's (2019) one, who also defends a view combining functionalism and primitive ontology. In Esfeld's account, the 3D primitive ontology is regarded as fundamental, and the wavefunction is to be regarded "as a dynamical parameter that is defined by its functional role for the evolution of matter" (p. 6253). Thus, his account is an *alternative* to WFR and is more akin to Allori (2021) – whereas Albert's one is based on WFR – and falls into the same category of the primitive ontology approaches introduced in sect. 2.

metaphysical issues in the next section. In particular, Ney has stressed that:

> There is no common 3D space for interparticle interactions, let alone interparticle distances. Each shadow lives in its own 3D subspace. For example, when $i = 1$ and $j = 2$, note the fact that the potential energy $V$ in [(22)] partly depends on the value $x_i - x_j$ or $x_1 - x_2$ which is simply the distance (in the x-dimension) between particles 1 and 2. In [(21)], $V$ depends on the value $x_{(3i-2)} - x_{(3j-2)}$ or $x_1 - x_4$ where this refers to the numerical difference between the first coordinate of the center of mass of the first shadow in its 3D subspace and the first coordinate of the center of mass of the second shadow in its own (different) 3D subspace. This isn't a distance; it's merely a difference between values. And so, there is no common 3D space in which the shadows may "coagulate" and make up tables and chairs. (Ney, 2021b, p. 216).

Ney claims that the kind of dynamics postulated by Albert cannot actually functionally realize a 3D ontology. If we cannot find a convincing way to say that the wavefunction of the theory can really enact 3D entities, then any functionalist version of the theory would be unviable. Thus, Albert's account has a serious challenge to face.[43] Indeed, it is so crucial to Ney (2015, 2021b) that this is the very problem that led her to abandon the functionalist approach to WFR she once defended (Ney (2012)).

In contrast, the account about the dynamics of the wavefunction that I have presented arguably avoid this objection. Ney's objection rests mainly on the fact that Albert merely postulates the two Hamiltonians and bases his proposal on an alleged formal similarity between the twos. On the other hand, the functionalist model I have proposed is closer to scientific practice and to the standard way of formulating non-relativistic quantum mechanics. In the limit cases where Ehrenfest theorem can be applied it follows naturally that the wavefunction behaves isomorphically to Newton's law, and there was nothing physically controversial about that model.

Summing up, in this section I have pointed out an important advantage of my account with respect to Albert's one, since the version of functionalism I have defended in section 3 is more detailed concerning the physics underlying the functionalisation process and avoids the challenge posed by Ney, which arguably arises due to the lack of formal details that affects the current literature about functionalism as applied to WFR. Furthermore, the next section shows that the functional reductionist approach defended here solves also several other problems which can be raised against WFR.

## 5. The virtues of functional reductionism

The first aspect on which Lewisian functional reduction is helpful is the clarification of the ontology of WFR. In fact, let's focus on Minimal Functionalism. That general framework does not tell us much about the nature of the non-fundamental ontology according to WFR. Indeed, claiming that 3D entities exist as long as there is something which plays the right functional role does not say much about their ontological status – besides that we should be realist about them – or about the exact metaphysical relation between the high-dimensional and low-dimensional entities. Are 3D entities identical with wavepackets? Do they supervene, or are determined by it? Do they inhabit the same ontological level of the wavefunction or not? Should we be reductionist about them or not? All those questions remain open, if one just consider Minimal Functionalism.

The topic is partially addressed in the literature. Albert (2013, 2015) claims that the fundamental 3N-dimensional ontology 'formally enacts' the elements composing the 3D ontology, and that the latter emerges from the former. Similarly, Ney (2012) talks about the 'realization' of the 3D ontology by the wavefunction within the functionalist account, while in Ney and Albert (2013) she sketches a reductionist view. Additionally, other authors like Chalmers (2021) and Rubenstein (2020, p. 13) describe the 3D ontology as non-fundamental and grounded in the fundamental high-dimensional level. Summing up, this highlights that no clear account seems to be available in the literature, since it's unclear how to combine all these notions together, and how to choose between them. Moreover, a generic appeal to emergence, reduction, realization or instantiation leaves unclear the exact ontological commitment of WFR with respect to the 3D ontology. Indeed, as stressed by Le Bihan (2018, 2019), depending on how one states the functionalist thesis, one can either maintain that (i) the 3D space is derivative; (ii) 3D states are identical with high-dimensional ones; (iii) be eliminativist about the 3D level.

In contrast, I argue that framing WFR in terms of functional reductionism gives us a clear principled way to address all those question. Indeed, having settled identity as the correct relation between the wavefunction and the 3D particles, the questions about the ontological commitments of WFR can be answered. Identity backs up a strongly reductive position concerning the nature of the 3D ontology, and this clearly supports the idea that there is only one level of reality involved, ontologically speaking. Moreover, the claim that wavepackets *are* 3D objects when they behave appropriately allow us to be realist about the 3D ontology. Eventually, the identity-based functionalism presented here stands in the middle between a derivative and an eliminativist position, by providing us with real but non-derivative 3D ontology. Moreover, *contra* Albert, the arguments in section 3.2 showed that identity relations logically follows from Minimal Functionalism, once one formalize more carefully (thanks to the Lewisian analysis) the Minimal Functionalist stance. Thus, one cannot maintain Minimal Functionalism and at the same time postulate an emergent or derivative 3D ontology, as he suggests.

Moving on, the functional reductionist framework presented here dissolves also several metaphysical issues that are linked to the relation between the high-dimensional and the low-dimensional realms. First, Maudlin (2019) has recently raised an objection related to the possibility of building arbitrary mappings between the high-dimensional and the 3D ontologies. Second, the relation between the levels has been dubbed as obscure or unscrutable (Maudlin, 2007 and Allori (2013)).[44] Third, the ability of the high-dimensional ontology to really *constitute* the 3D particles has been questioned (Ney (2015, p. 3117)).

I believe that this general scepticism against WFR is ultimately a consequence of the lack of a clear account about the ontological relation between the high and the low-dimensional ontology. As I have highlighted at the beginning of this section, a generic appeal to functionalism alone is not sufficient to settle all the ontological issues: the idea that 3D objects exist insofar as there is something which plays the right causal role leaves open the question about the metaphysical relation between the 3D objects and their realizers. This fact, combined with the generic appeal to notions such as realization, instantiation or emergence, could give us the impression that the low-dimensional ontology is somehow *generated* by the high-dimensional one. That is, it gives the impression that there is some kind of *gap* between the two. Furthermore, notions like realization or instantiation do not possess a clear and univocal meaning within the philosophical literature, and the way they are used within the debate on WFR make them appear more as proxy terms than as precise relations.[45] Thus, this situation can lead one to think that we need a further explanation – on top of functionalism – to explain why and how

---

[43] Albert develops a very similar account for Bohmian mechanics – proviso the addition of the marvelous point to the fundamental ontology – and the same concerns arguably apply there as well.

[44] Rubenstein (2020, sect. 4) provides an excellent description of this issue.
[45] The same issue is present in the literature about spacetime emergence in quantum gravity, where 'emergence' is used somehow as a proxy term in need of clarification (see Wüthrich (2017), 318).

the two ontologies are connected. And this is where the functional reductionism helps.

As I have pointed out, this framework gives us a clear account about the relation between the elements of the two ontologies – i.e. identity – and provides a reductive *explanation* of how that identity obtains. In other words, that relation is not a brute fact. Instead, it follows deductively, and should not be deemed as obscure, arbitrary or unscrutable.[46] On the other hand, if one wants to contrast this explanation, one has the onus of finding out some functional role associated with 3D objects which cannot be played by the high-dimensional ontology, or some intrinsic feature of 3D objects which cannot be functionalized. If one does not succeed, then the Lewisian analysis will go through: no gaps are left. Thus, we can say the 3D ontology is nothing over and above the wavefunction: once the high-dimensional ontology can play the right roles, nothing *more* has to be added to secure the constitution of the 3D ontology.[47] In the philosophy of mind, the non-functionalizable intrinsic features are the so-called qualia, but in our context it is far from clear that any similar feature is in place.[48]

Finally, I think that ruling out ontological and explanatory gaps is particularly helpful to address an important challenge raised by Maudlin (2019, p. 126), concerning the possibility of multiple mappings between the high and the low-dimensional ontologies. Following Rubenstein (2020), we can call it the *displaced ghosts objection*. Recall that within Albert's proposal (sect. 4) the wavefunction enacts what he calls 3D 'shadows'. Maudlin argues that, in addition to Albert's mappings from the 3N-dimensional to the 3D space (the shadows), there are also 'ghost-worlds': if you take the global translation 'three feet to the North', you get a world with the same geometrical and dynamical structure of the one you started with. Thus, this ghost-world should be equally regarded as real. What prevents those 'ghost-worlds' from existing? Actually, functional reductionism can help us to dissolve this puzzle, since it allows to close any alleged gap between the high-dimensional and the low-dimensional levels. In fact, within this account, we do not have to postulate the existence of a distinct derivative 3D ontology. Instead, we have identity relations between the two ontologies which follows deductively, and are not postulated *by fiat*, and so no gap is left. Thus, since there's no ontological distinction between 3D and high-dimensional entities, it seems difficult to think how the latter entities can ground alternative and distinct ghost-worlds. After all, identity is transitive.[49]

Having addressed these inter-related objections, I want to focus on two further final pros of my account. First, recall the first advantage I mentioned in the introduction. I claimed that the functional reductionist version of WFR dissolves some issues which have been recently raised against Albert's functionalist approach to WFR, especially by Ney (2021b). One issue is the one discussed in section 4, which I tackled by providing a more accurate model concerning the possible classical behaviour of the wavefunction. Another objection that Ney raises against Albert's functionalism is the following:

> Albert's approach […] ties the wavefunction's enactment of a 3D ontology to its approximation of classical behavior. This raises the question of how the wavefunction realist might recover those macrosystems exhibiting nonclassical, that is, distinctively quantum behavior. […] It would be nice to at least have available a way to recover three-dimensionality that does not assume classicality. (Ney, 2021b, p. 219).

This objection is formulated against Albert's proposal *qua* functionalist account, and thus it could be supposed to concern similarly also my proposal. The crucial problem, according to her, is the fact that three-dimensionality is tied to classical behaviour within functionalism. However, given what I have said at the end of the last section, I think it is clear that this objection does not have much force. Indeed, it could be even said that she is begging the question, given that she is assuming here that we should account for *non-classical 3D* objects. In particular, she wonders how the functionalist version of WFR can recover macro-system which behave quantum-mechanically. However, as I argued, the functional reductionist approach to WFR can actually easily account for macro-systems which behaves non-classically. Such a system, in the ontology of WFR, would simply correspond to a wavefunction evolving in configuration space. Indeed, there is no need to say that such object is 3D, if we are committed to a fundamentally high-dimensional ontology. A 3D system, under this understanding of the ontology of quantum mechanics, corresponds to a specific state of the wavefunction when the wavefunction behaves appropriately. It would not make any sense, therefore, to decouple classicality and three-dimensionality within this account.

Finally, to conclude, I want to discuss briefly one last advantage of the version of WFR defended in the paper (the fourth point in section 1.2). Let's recall that one alleged advantage of WFR is being an account about the ontology of non-relativistic quantum mechanics which follows in a straightforward way from the mathematical formulation of the theory (under a few assumptions). Additionally, I mentioned that WFR can preserve both fundamental separability and locality. On the other hand, primitive ontology approaches do not preserve both these latter features. Moreover, those accounts do not simply read the ontology of quantum mechanics directly from the plain formalism. Instead, they postulate additional ontology at the fundamental level (which is 3D, according to them). Ultimately, they do so because they think that starting from a 3D ontology is more intuitive and gives a conceptually clearer ontology – made of microscopically 3D building blocks on which macro-objects can be constructed.

What I want to stress here is that the functional reductionist version of WFR is a perfect halfway house between standard WFR (such as Albert's)[50] and the primitive ontology view. That is, this particular version of WFR can be argued to be the best of both worlds, combining the conceptual clarity of the latter account with the pros of the former one. Thanks to functional reduction, we are not committed to a layered view of reality, where the high-dimensional ontology lies at a lower level and is connected in some ambiguous way to the emergent 3D level. Instead, the wavefunction (or more generally the high-dimensional ontology) *is* simply 3D, in the appropriate contexts. Thus, we can have 3D entities at the fundamental level, in those specific situations where classical behaviour is displayed – and so it's not clear anymore what kind of conceptual advantage would the primitive ontology view have over WFR.

## 6. Conclusion: what Wave Function Realism can teach us

Concluding, I want to stress the take-home messages of this essay, beyond the restricted debate about WFR. I think that the present discussion about WFR as a case study and the discussion about functional reductionism can be useful on two broader levels.

First, a methodological lesson. That is, I have proven how improving the mathematical and logical details of the discussion brings key advantages. On the one hand, in section 3.1 I have presented a clearer account about the physics underlying the functionalist account of WFR, by discussing the functionalisation in the context of some real world cases. I have then argued that this functionalisation model (based on Ehrenfest's theorem and thus on an uncontroversial piece of standard quantum mechanics which is relevant in scientific practice) fares better than Albert's account which is roughly sketched on the alleged similarity between two postulated Hamiltonians. On the other hand, in sections 3.2 and 3.3 I have shown how we can formalize in a logically preciser way the informal functionalist intuitions behind the Minimal Functionalist approach to WFR. I have presented how, following this formalization, a form of functional reductionism follows naturally from the abstract functionalist account – and this leads to important consequences (and advantages). Summing up, I have showed the benefits of bringing the discussion about functionalism in philosophy of science to a higher level of formal precision.

---

[46] Chalmers (2021) defends a similar view.
[47] A similar claim is defended by Wüthrich (2019, p. 324) and by Lam and Wüthrich (2018, p. 10), in their defence of a functionalist account for spacetime in quantum gravity.
[48] See Chalmers (2012, 2021).
[49] Chen (2017) suggests a similar modification of the Minimal Functionalist approach, to avoid the displaced ghosts objection (although he discards this strategy for other reasons). He grants that if we implement the functionalist approach with identity mappings the problem is dissolved, since "identity is a strict relation that is not preserved by arbitrary mappings" (p. 350).

[50] And, in general, any form of WFR discussed in the literature, including Ney (2021b) non-functionalist WFR.

This brings us to the second bunch of lessons we can learn from our case study. In fact, some of the advantages of functional reductionism that I have discussed in section 5 can be reasonably expected to be widely applicable beyond WFR. For instance, the fact that functional reductionism allows us to bring together the advantages of the fundamental ontology of WFR with the conceptual clarity of a 3D ontology can be transferable to other cases. Moreover, the facts that functional reductionism gives us a clear ontology about WFR, which was lacking in the Minimal Functionalist approach as well as in Albert's account, and that it closes the gap between the bottom and the upper levels, are very general features that we can apply to other contexts in science. In the case of WFR, functional reductionism helps us to tackle some important objections which have been much discussed: now that the discussion has been flashed out, I hope that the arguments of the last section can block the insurgence of the same kind of objections in other debates about functionalism in science. In general, one of my aims has been to pave the way for the application of the functional reductionist approach – as presented here – to other contexts such as quantum gravity, Wallace's Everettian account or thermodynamics, where functionalism has been recently employed.

### Acknowledgments

I am very grateful to Karim Thébault, James Ladyman, Alexander Franklin, Eleanor Knox, Christian Wüthrich, Baptiste Le Bihan, Tim Maudlin and Claudio Calosi for their helpful comments and discussions on several versions of this paper. I also would like to thank the anonymous referees for their feedback, as well as the audiences at the 2021 annual conference of the British Society for the Philosophy of Science, the 6th annual conference of the Society for the Metaphysics of Science, the 8th biennial meeting of the European Philosophy of Science Association, and the Geneva Symmetry Group seminars series. This research has been supported and funded by SWW DTP (AHRC).